\documentclass[aps, prl, reprint, superscriptaddress]{revtex4-2}
\pdfoutput=1
\usepackage{braket,dsfont,subfigure,amsfonts,amssymb,bm,graphicx,float,amsmath,amsthm,txfonts}
\usepackage[colorlinks]{hyperref}
\usepackage{orcidlink}

\begin{document}
\title{Unveiling Stripe-shaped Charge Density Modulations in Doped Mott Insulators}

\author{Ning Xia}
\affiliation{State Key Laboratory of Low Dimensional Quantum Physics and Department of Physics, Tsinghua University, Beijing 100084, China}
\author{Yuchen Guo~\orcidlink{0000-0002-4901-2737}}
\affiliation{State Key Laboratory of Low Dimensional Quantum Physics and Department of Physics, Tsinghua University, Beijing 100084, China}
\author{Shuo Yang~\orcidlink{0000-0001-9733-8566}}
\email{shuoyang@tsinghua.edu.cn}
\affiliation{State Key Laboratory of Low Dimensional Quantum Physics and Department of Physics, Tsinghua University, Beijing 100084, China}
\affiliation{Frontier Science Center for Quantum Information, Beijing 100084, China}
\affiliation{Hefei National Laboratory, Hefei 230088, China}

\begin{abstract}
Inspired by recent experimental findings, we investigate various scenarios of the doped Hubbard model with impurity potentials. 
We calculate the lattice Green's function in a finite-size cluster and then map it to the continuum real space, which allows for a direct comparison with scanning-tunneling-microscopy measurements on the local density of states. 
Our simulations successfully reproduce experimental data, including the characteristic stripe- and ladder-shaped structures observed in cuprate systems. 
Moreover, our results establish a connection between previous numerical findings on stripe-ordered ground states and experimental observations, thus providing new insights into microscopic mechanisms of the Mott insulator to superconductor transition in cuprates.
\end{abstract}
\maketitle

\emph{Introduction.---} 
Understanding the high-temperature superconductivity in cuprates remains a fundamental challenge in condensed matter physics. 
Materials such as $\rm{YBa_2Cu_3O_{6+y}}$ (YBCO), $\rm{Bi_2Sr_{2-x}La_xCuO_{6+y}}$ (Bi2201), and $\rm{Ca_{1-x}Na_xCuO_2Cl_2}$ (NCCOC) exhibit a unified phase diagram.
In the undoped region, they are Mott insulators with long-range antiferromagnetic order.
Upon hole doping, they undergo a phase transition to become superconductors.
This shared behavior suggests a common underlying mechanism within the cuprate family~\cite{dopingmott,remarkable}. 

Although phase fluctuation theory~\cite{dopingmott,emery1995importance} provides a framework for understanding the phase diagram of cuprates, the microscopic mechanisms driving the transition from Mott insulator to superconductor are not fully understood. 
A recent scanning tunneling microscopy (STM) experiment~\cite{ye2023emergence} on underdoped Bi2201 probes this critical region and observes modulations in local density of states (LDOS) near the Fermi energy. 
In the insulating sample with a hole concentration of $p=0.08$, there is a spatial phase separation~\cite{emergy_kivelson_lin_phase_separ,Dagotto_Review} between the hole-free antiferromagnetic phase and the hole-rich phase. 
The spatial variations within the hole-rich region are highly analogous to those observed in the superconducting sample with $p=0.11$.
Specifically, the doped holes self-organize in short-range checkerboard order~\cite{checkerboard,cai2016visualizing} with a wavelength of approximately $4a_0$ (where $a_0$ is the lattice constant of the $\rm{CuO_2}$ plane), and each plaquette in the checkerboard displays an internal stripe pattern~\cite{haihu,zaanen2023high}. 
The primary difference between samples lies in the spatial occupation of the plaquettes. 
These findings suggest that plaquettes with internal stripe-shaped patterns are essential elementary building blocks. 
Moreover, even finer spatial structures including ladder- and clover-shaped patterns are observed at higher energy in lightly doped NCCOC~\cite{ye2023}, which are attributed to molecular orbitals embedded in a Mott insulator~\cite{haiwei}.
However, systematic investigations are necessary to corroborate their intuitive understanding.

From a theoretical perspective, the Hubbard model and $t$-$J$ model are considered minimal models that qualitatively capture the essence of cuprates~\cite{Hubbard,chen1,chen2,chen3,anderson2004,White,Grilli,emergy_kivelson_lin_phase_separ,absence,xustripe,QinHubbard,whitesc,whitephasesep,Wengtwohole,luxin,lu2023signstructuresquarelatticettprimej,chen2023dwavepairdensitywavesuperconductivitysquarelattice,zhao2023distinctquasiparticleexcitationssingleparticle,phase_string,phase_string_general,xu2024globalphasediagramdoped,yue2024pseudogapphasefluctuatingpair,Dagotto_Review}.
At low doping levels, the ground state exhibits various charge orders~\cite{emergy_kivelson_lin_phase_separ,whitestripe,whitephasesep}, including phase separation, uniform, and stripe configurations with energy close to each other and thus competitive.  
On the other hand, inhomogeneous impurity potentials of dopants can readily destroy the uniform state in realistic materials, thereby favoring phase separation. 
To further explain the spatial structures observed in experiments~\cite{ye2023,ye2023emergence,checkerboard}, previous studies investigate the ground state of these models in a finite-size cluster. 
For example, simulations using the density matrix renormalization group (DMRG) on the $t$-$J$ model~\cite{White} show that external potential fields can induce charge density modulations in the stripe state~\cite{corbozstripe,whitestripe,whitetj,xustripe}, resembling the checkerboard state. 
With the Gutzwiller approximation, the Hubbard model also produces a checkerboard-like pattern~\cite{Grilli}.
However, differential conductance and LDOS measured with STM are related to the retarded Green’s function in the continuum real space according to the linear response theory~\cite{stm}, which is not directly comparable to those numerically simulated ground state properties such as charge and spin distributions in a lattice system.

In this paper, inspired by experiments and the concept of molecular orbitals in Mott insulators~\cite{ye2023emergence,haiwei,ye2023}, we investigate the stripe-ladder pattern of the doped Hubbard model with impurity potentials. 
We first employ the Chebyshev matrix product state (CheMPS) approach~\cite{Che1,Che2,Che3} to compute the retarded lattice Green's function. 
Subsequently, we apply a basis transformation~\cite{Interpretation1,Interpretation2} to effectively map the Green's function to the continuum real space for direct comparison with experimental observations~\cite{ye2023}. 
By tuning the impurity potentials in our simulations, we successfully reproduce the experimental data~\cite{ye2023}, particularly the newly discovered stripe-ladder pattern.
This outcome offers new insights into the formation of stripe orders and demonstrates the potential of our approach to explain the complex behavior of cuprates.

\begin{figure}[tbp]
    \includegraphics[width=0.9\linewidth]{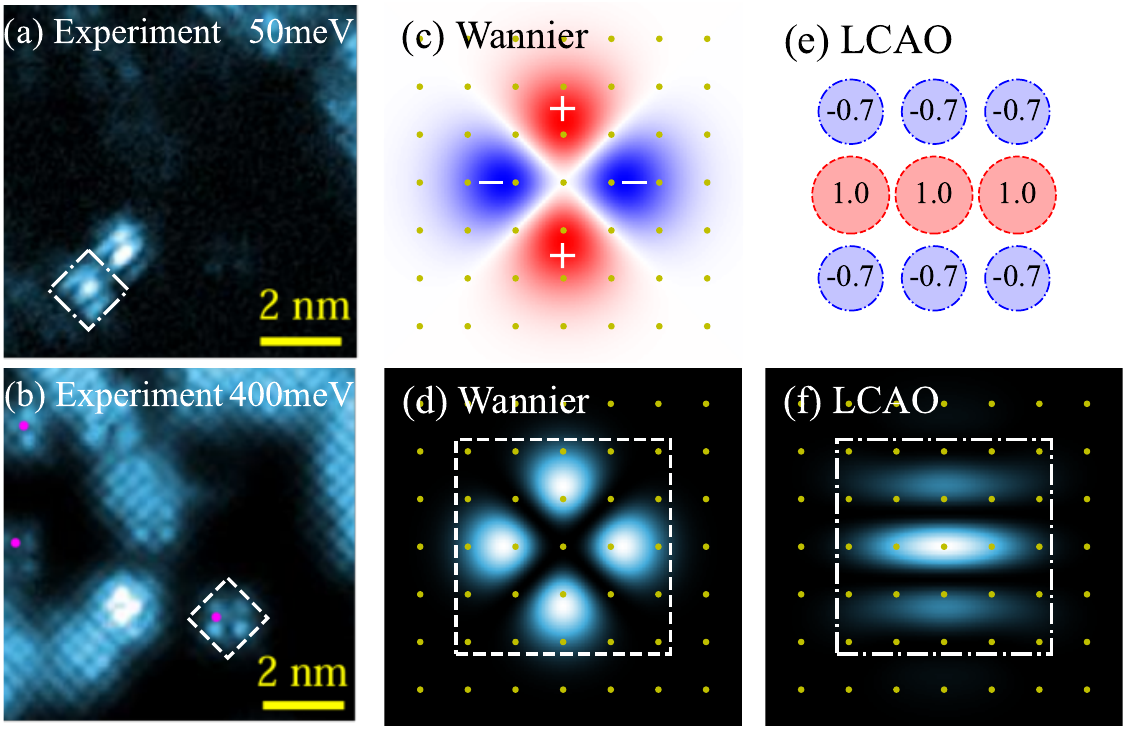}
    \caption{(Color Online)
    (a-b) Experimental results of LDOS at $50\textrm{meV}$ (a) and $400\textrm{meV}$ (b), respectively, taken from Ref.~\cite{ye2023}. 
    (c-d) Density maps of the approximated Wannier function $w_0(x,y)$ (c) and $|w_0(x,y)|^2$ (d) at the surface exposed to the STM tip. 
    Yellow dots represent Cu sites. 
    (e) The coefficients for LCAO. 
    (f) The density distribution of the LCAO wavefunction.
    }\label{fig:1}
\end{figure}

\emph{Model and method.---} 
We begin by interpreting the experimental data from Ref.~\cite{ye2023}. 
As illustrated in Fig.~\ref{fig:1}(a, b), the underdoped sample with a hole concentration of $p=0.03$ exhibits phase separation into dark antiferromagnetic regions and bright hole-rich regions.
In the hole-rich region of Fig.~\ref{fig:1}(a), two similar plaquettes emerge with an internal stripe-shaped pattern. 
In Fig.~\ref{fig:1}(b), a ladder-shaped pattern emerges in the same area.
Moreover, a four-lobe clover-shaped pattern appears near the Na dopant marked by the magenta dot. 
In particular, this clover-shaped pattern is centered on the Cu sites, rather than the Na dopant~\cite{ye2023}. 

Based on these observations, it is proposed that the clover-shaped pattern reflects the Wannier function of a single-hole state. 
For example, calculations in density functional theory (DFT)~\cite{Interpretation1,Interpretation2,clover} show that the planar structure of the Wannier function exposed to the STM tip exhibits a consistent four-lobe-clover shape for different cuprate materials, regardless of the specific details near the $\rm{CuO_2}$ plane. 
Moreover, the clover extends over neighboring $\rm{Cu}$ sites due to the spacing between the STM tip and the sample surface~\cite{clover}, in agreement with experimental observations~\cite{ye2023}.
It also exhibits similarities to localized Zhang-Rice singlets, whose matrix element effect in STM tunneling leads to a four-lobe-clover LDOS distribution~\cite{zhangrice,li2024boundstatesdopedcharge,imp2}.
To approximate this four-lobe pattern, we use the symmetrized combination of $p$-orbital wave functions~\cite{zhangrice}. 
This choice is consistent with previous research on the sign structure of wave functions~\cite{Interpretation1,Interpretation2,clover}, and does not induce qualitative differences in our final results compared to the $d$-orbital wave functions.
The density map of this approximated Wannier function $w_0(x,y)$ is shown in Fig.~\ref{fig:1}(c) with $d_{x^2-y^2}$ symmetry, whose density distribution $|w_0(x,y)|^2$ is further illustrated in Fig.~\ref{fig:1}(d), displaying a four-lobe clover shape with its size adjusted to compare with experiments. 
Following the molecular orbital proposal in doped Mott insulators~\cite{ye2023}, we consider a linear combination of Wannier functions at different sites in analogy to the linear combination of atomic orbitals (LCAO)~\cite{hoffmann,hoffmannrmp}, with coefficients shown in Fig.~\ref{fig:1}(e).
The resulting molecular orbital has a stripe-shaped distribution (Fig.~\ref{fig:1}(f)) similar to that observed in experiments.

In short, the LCAO argument~\cite{hoffmann,hoffmannrmp} provides a preliminary understanding of the stripe-shaped pattern in experiments. 
The impurity potential introduced by the Na dopants creates a region that attracts holes, allowing the Wannier functions inside this region to play dominant roles in LCAO.
However, systematic many-body numerical simulations instead of single-particle arguments are required to verify the emergence of stripe-shaped molecular orbitals.
Therefore, we continue to consider the Hubbard model defined on a finite cluster
\begin{equation}
\begin{aligned}
H=&-t_0\sum_{\langle ij\rangle,\sigma}c_{i\sigma}^\dagger c_{j\sigma} -t_1\sum_{\langle \langle ij\rangle \rangle,\sigma}c_{i\sigma}^\dagger c_{j\sigma}-\mu\sum_i n_i\\
&+U\sum_i(n_{i\uparrow}-\frac{1}{2})(n_{i\downarrow}-\frac{1}{2}) +\sum_i V_i n_i,
\end{aligned}
\end{equation}
where we set typical parameters $t_0=1.0$, $t_1=-0.3$, and $U=12$~\cite{KuDFTpara,Para2,Grilli,QinHubbard}. 
The number of holes is controlled using $U(1)$ symmetry~\cite{U1symm,itensor}. 
The chemical potential $\mu$ is tuned to ensure the ground state with the desired hole number being the true ground state of the entire system~\cite{absence,disorder,disorder2,percolation}. 
Local potential fields $V_i$ are applied to certain lattice sites to simulate impurity effects.

To solve this model and reproduce the experimental data, we first implement the standard DMRG procedure to obtain the ground state $|\Omega\rangle$ with energy $E_0$ and hole number $N_0$. 
We then apply the CheMPS method~\cite{Che1,Che2,Che3,SeeSM} to compute the retarded lattice Green's function $G_{ij,\sigma}^R(\omega)=G^+_{ij,\sigma}(\omega)-G^-_{ij,\sigma}(\omega)$.
Here,
\begin{equation}
\begin{aligned}
    &G^+_{ij,\sigma}(\omega) = \langle\Omega|c_{i\sigma} \frac{1}{(\omega+i0^+)+(E_0-H)}c_{j\sigma}^\dagger|\Omega\rangle,\\
    &G^-_{ij,\sigma}(\omega) = \langle\Omega|c_{j\sigma}^\dagger\frac{1}{-(\omega+i0^+)+(E_0-H)}c_{i\sigma}|\Omega\rangle.
\end{aligned}
\end{equation}
After that, we convert the retarded lattice Green's function to a continuum real-space representation by~\cite{Interpretation1,Interpretation2}
\begin{equation}
G^R_\sigma(\bm{r},\bm{r}',\omega)=\sum_{ij}w_i(\bm{r})w_j^*(\bm{r}')G_{ij,\sigma}^R(\omega),
\end{equation}
where $w_i(\bm{r})$ is the Wannier function centered at the $i$-th Cu site. 
The LDOS $\rho(\bm{r},E)=-\sum_\sigma\textrm{Im}G_\sigma^R(\bm{r},\bm{r},E)/\pi$ is proptional to the STM differential conductance $\textrm{d}I/\textrm{d}V$, allowing for direct comparisons. 

Although the results presented in this manuscript are calculated using a finite cluster with open boundary conditions, we also use a hybrid approach~\cite{zhao2024chebyshevpseudositematrixproduct} combining CheMPS~\cite{Che3,Che1,Che2} and cluster perturbation theory (CPT)~\cite{CPT1,CPT2,Pavarini_2015,realspaceCPT,Lijianxin} to study larger systems (see Supplemental Material~\cite{SeeSM}). The main conclusions remain qualitatively unchanged, suggesting that the physical phenomena observed in the finite cluster are robust and not artifacts of the finite-size simulation.

\begin{figure*}[tbp]
    \includegraphics[width=0.8\linewidth]{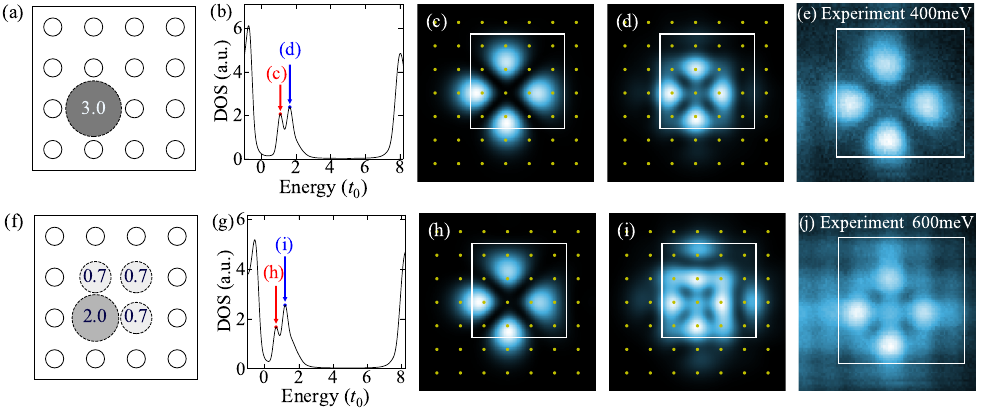}
    \caption{(Color Online) 
    (a-d) The first one-hole case with chemical potential $\mu=-3.7t_0$. 
    (a) Model defined on a $4\times 4$ cluster, with circles representing Cu sites. 
    Local potential fields $V_i$ are applied in gray circles, with radii proportional to $|V_i|$. 
    (b) DOS for the first one-hole case. 
    (c-d) The LDOS at two peaks in (b). 
    The sites inside the white box correspond to those calculated during the simulation. 
    (f-i) Results for the second one-hole case with chemical potential $\mu=-3.7t_0$. 
    (e, j) Experimental LDOS for lower energy at $400\rm{meV}$ (e) and for higher energy at $600\rm{meV}$ (j), taken from Ref.~\cite{ye2023}.
    }\label{fig:2}
\end{figure*}

\emph{Numerical results.---} 
We perform simulations of doped holes in finite clusters, including three scenarios based on the number of doped holes: the one-hole, two-hole and four-hole cases. 
In the one-hole case, we first apply a strong potential field to a single lattice site to pin the doped hole for a $4\times 4$ cluster as shown in Fig.~\ref{fig:2}(a), whose results are illustrated in Fig.~\ref{fig:2}(b-d).
The density of states (DOS) is shown in Fig.~\ref{fig:2}(b), where the lower Hubbard band lies below energy $0$ and the upper Hubbard band lies above energy $7t_0$.
Doping-induced in-gap states appear between these two bands, with two distinct peaks marked by red and blue arrows. 
These in-gap peaks reflect the spectral weight transfer phenomena~\cite{anderson2kf,swt,SWT2,SWT3,SWTRMP,cai2016visualizing,RUAN20161826}.
In the localized limit, doping a hole leaves behind an empty site where an electron can be added in two possible configurations (spin-up and spin-down). 
In our model, the added electron must overcome the impurity potential field while interacting with the surrounding antiferromagnetic background, leading to the splitting into two peaks~\cite{SeeSM}.
The real space structures of these peaks are visualized in Fig.~\ref{fig:2}(c) and Fig.~\ref{fig:2}(d), respectively. 
The LDOS for the lower energy peak exhibits a four-lobe clover pattern almost identical to the shape of the Wannier function, centered at the Cu site where the potential field is applied (Fig.~\ref{fig:2}(c)), consistent with the experimental results in Fig.~\ref{fig:2}(e).
In contrast, the LDOS for the higher energy peak shows a halved clover pattern in Fig.~\ref{fig:2}(d).
These features are qualitatively comparable with Fig.~\ref{fig:2}(j), while the doped hole spreads more extensively at higher energy in the experimental results, particularly within and outside the clover-shaped pattern.

To better simulate the higher-energy structure, we consider an additional set of parameters in Fig.~\ref{fig:2}(f). 
Ideally, the potential field generated by the Na dopant is symmetric with respect to the nearest-neighbor Cu sites~\cite{imp,imp2} (Fig.~S1(a) in Supplemental Material~\cite{SeeSM}), while experiments~\cite{ye2023} suggest that the nearby dopants and defects can break this symmetry.
Taking this into account, we apply potential fields $V_i$ to four nearby lattice sites, with one dominant over others.
The DOS of the in-gap state in Fig.~\ref{fig:2}(g) still shows two characteristic peaks.
For the lower energy peak shown in Fig.~\ref{fig:2}(h), it exhibits a clover-shaped distribution centered at the Cu site, where the potential field is strongest. 
For the higher energy peak in Fig.~\ref{fig:2}(i), four sites with higher hole density display a halved clover shape similar to that in Fig.~\ref{fig:2}(d).
However, the density distribution here is more extensive and thus better reproduces the experimental results in Fig.~\ref{fig:2}(j), suggesting that the effective impurity potentials in experiments may have a distribution similar to that in Fig.~\ref{fig:2}(f).

\begin{figure}[tbp]
    \includegraphics[width=0.9\linewidth]{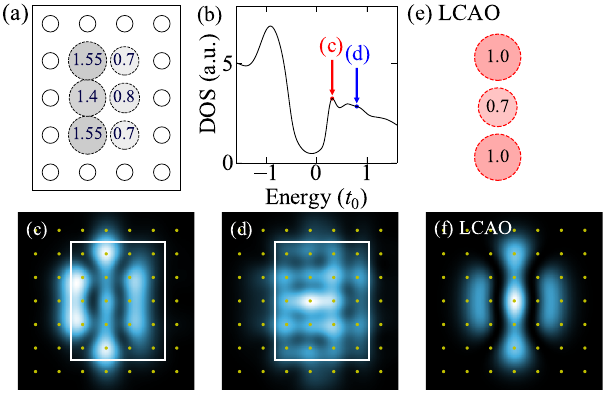}
    \caption{(Color Online) 
    Two-hole case with chemical potential $\mu=-3.7t_0$. 
    (a) Model defined on a $5\times 4$ cluster. 
    (b) DOS of the two-hole case. 
    (c-d) LDOS at energies highlighted by the red and blue arrows in (b). 
    (e) LCAO coefficients. 
    (f) Density distribution of the LCAO result.
    }\label{fig:3}
\end{figure}

Moving on to the two-hole case, we aim to verify that the observed stripe-shaped pattern reflects the molecular orbital embedded in the Mott insulator.
We consider a $5\times 4$ cluster shown in Fig.~\ref{fig:3}(a), with impurity potential fields $V_i$ applied to the inner $3\times 2$ region and dominant on one side, mimicking the effects of other impurities. 
The DOS in Fig.~\ref{fig:3}(b) shows a gap structure near the Fermi energy, with the part below $0$ related to the lower Hubbard band and the part above $0$ associated with the in-gap state.
The peak at energy $0.32t_0$ as a characteristic lower-energy state and a higher-energy state at $0.8t_0$ for reference are marked by red and blue arrows.
The lower-energy LDOS in Fig.~\ref{fig:3}(c) shows a three-stripe pattern, resembling an alternative LCAO outcome depicted in Fig.~\ref{fig:3}(e, f), where the Wannier functions form an in-phase superposition analogous to that of a bonding molecular orbital.
For the higher-energy state in Fig.~\ref{fig:3}(d), we see a ladder-shaped distribution of LDOS perpendicular to the lower-energy stripe-shaped distribution, which corresponds to an antibonding molecular orbital and agrees with the experimental results.

\begin{figure}[tbp]
    \includegraphics[width=\linewidth]{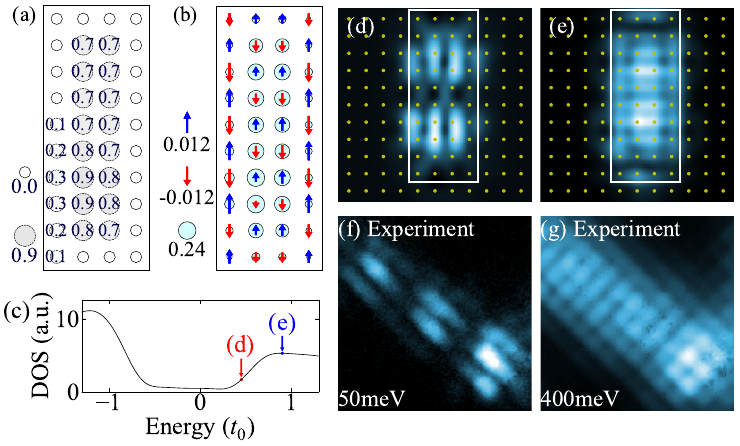}
    \caption{(Color Online) 
    Four-hole case with chemical potential $\mu=-3.7t_0$. 
    (a) Model defined on a $10\times 4$ cluster.
    (b) The hole densities and the magnetic moments of the ground state. 
    The circle and arrows outside the box illustrate the maximums of the hole densities and the magnetic moments, respectively. 
    (c) DOS of the four-hole case. 
    (d-e) LDOS at energies indicated by the arrows in (c). 
    (f-g) Experimental LDOS for lower energy at $50\rm{meV}$ (f) and for higher energy at $400\rm{meV}$ (g), taken from Ref.~\cite{ye2023}.
    }\label{fig:4}
\end{figure}

We successfully reproduce the stripe-ladder pattern in the two-hole case (more results in Fig.~S2~\cite{SeeSM}).
However, to demonstrate the checkerboard structure with an internal stripe-ladder pattern, we need to consider a minimal four-hole case. 
In this scenario, a larger number of dopants will induce smoother local potential fields.
To simulate this, we trap four holes inside a $10\times 4$ cluster with broad potential fields, as shown in Fig.~\ref{fig:4}(a).
We first calculate the hole densities $1-\langle (n_{i\uparrow}+n_{i\downarrow})\rangle$ and the magnetic moments $\frac{1}{2}\langle (n_{i\uparrow}-n_{i\downarrow})\rangle$ for our ground state in Fig.~\ref{fig:4}(b).
The hole density is concentrated mainly in the central $8\times 2$ region with slight modulations, and this region also forms a domain wall of the surrounding antiferromagnetism~\cite{zaanen_domain,MACHIDA1989192charge,machida_domain,zhangQCS}.
It indicates that the ground state is indeed a stripe state, consistent with previous numerical results for the $t$-$J$ model~\cite{White}.

In addition, the stripe state is deemed relevant to the checkerboard structure observed in experiments~\cite{White}.
By simulating spectral functions, we better demonstrate this correspondence through the characteristic features of excited states that exhibit checkerboard plaquettes with internal stripe-ladder patterns. 
In particular, Fig.~\ref{fig:4}(c) shows the DOS around the Fermi energy, which displays a U-shaped gap structure. 
Due to the large degrees of freedom in this system, characteristic peaks are no longer clearly visible.
Therefore, we highlight two typical energies at $0.45t_0$ and $0.9t_0$, marked with red and blue arrows.
We plot the LDOS for these two energies in Fig.~\ref{fig:4}(d) and Fig.~\ref{fig:4}(e), respectively.
At lower energy, we find that the DOS exhibits two plaquettes with three stripes inside each plaquette.
This pattern qualitatively replicates the experimental data shown in Fig.~\ref{fig:4}(f). 
Notably, the lengths and intensities of these three stripes are roughly equivalent. 
This result is in contrast to the two-hole case depicted in Fig.~\ref{fig:3}(c), where the central stripe is longer and less intense.
The similarity between our simulated data and the experimental results suggests that in the four-hole case, the pairs of holes in two plaquettes interact with each other, resulting in a shorter and thicker central stripe.
This result highlights the role played by the interaction between holes in forming such a many-body pattern.
At higher energy, we observe an LDOS pattern that exhibits a ladder-shaped distribution, as shown in Fig.~\ref{fig:4}(e).
This qualitative agreement with the experimental data (Fig.~\ref{fig:4}(g)) indicates that our simulations capture the essence of experimental phenomena (more results in Fig. S3~\cite{SeeSM}).
These findings demonstrate the ability of our simulations to model and predict the behavior of this complex system, providing valuable insights into the underlying physics.

\emph{Conclusion and discussion.---}
In summary, we employ the CheMPS method to simulate the retarded lattice Green's functions for various hole-doping cases in the doped Hubbard model with impurity potential fields.
Our calculations enable the projection of the LDOS onto the continuum real space, allowing for a direct and accurate comparison with the STM experiments. 
We find that the single-hole state exhibits a four-lobe clover pattern in its lowest in-gap state. 
In contrast, the two-hole state forms a stripe-ladder pattern, which is strikingly similar to the bonding and antibonding molecular orbitals formed by the dimerization of two holes.
This result is consistent with experimental observations. 
Notably, the four-hole state shows two distinct LDOS patterns depending on the energy range.
At lower energy, we observe two stripe-shaped plaquettes that are characteristic of the interaction and correlation between the pair of holes.
At higher energy, a ladder-shaped structure emerges, in agreement with previous experimental studies~\cite{ye2023}. 
Our calculations also reveal the hole densities and magnetic moments of the ground state, which exhibit a slightly modulated stripe order. 
This result is consistent with previous numerical calculations using the $t-J$ model~\cite{White}, further confirming the validity of our simulations.

As proposed in Ref.~\cite{ye2023,haiwei}, the stripe-ladder patterns of the doped holes exhibit striking similarities to molecular orbitals embedded in antiferromagnetic Mott insulators. 
Our simulations not only provide supportive evidence for this proposal, but also offer a unique perspective on the many-body physics involved.
Specifically, our results imply that doped holes occupying such molecular orbitals can give rise to a many-body ground state with stripe order~\cite{White,zaanen_domain}. 
A particularly intriguing feature of our finding is the presence of a stripe-shaped plaquette containing two holes, suggesting a possible pairing structure in the ground state.
This result resonates with recent theoretical studies that propose paired twisted holes as a mechanism for charge order in doped Mott insulators~\cite{Wengtwohole,zhao2023distinctquasiparticleexcitationssingleparticle,phase_string,phase_string_general}. 
To gain a deeper understanding of this phenomenon, more research is needed to explore the properties of excitations in doped Mott insulators. 

Our study successfully bridges the gap between numerical simulations and experimental observations, providing new insights into the intricate interplay between antiferromagnetism and charge order.
We believe that our results will provide inspiration for future related research and shed light on the complex physics of strongly correlated systems, ultimately contributing to a better understanding of doped Mott insulators.

\emph{Note added.---}
As we finalize this manuscript, we become aware of an independent study focusing on the three-band Hubbard model~\cite{li2024boundstatesdopedcharge}.
Their investigation reveals that the in-gap state exhibits a four-lobe clover-shaped pattern in LDOS, which they attribute to the localized Zhang-Rice singlet.
This finding is consistent with our choice of basis for the one-band Hubbard model and our interpretation for a localized single hole.

\begin{acknowledgments}
\emph{Acknowledgement.---} 
We thank Shusen Ye and Yayu Wang for helpful discussions and providing STM data. 
We also thank Zheng-Yu Weng for useful suggestions.
This work is supported by the National Natural Science Foundation of China (NSFC) (Grant No. 12174214 and No. 12475022) and the Innovation Program for Quantum Science and Technology (Grant No. 2021ZD0302100).
\end{acknowledgments}

\bibliography{reference}

\clearpage
\appendix
\newpage
\onecolumngrid
\renewcommand{\theequation}{S\arabic{equation}} \setcounter{equation}{0}
\renewcommand{\thefigure}{S\arabic{figure}} \setcounter{figure}{0}

\begin{center}
\large{\textbf{Supplemental Material}}
\end{center}

\renewcommand{\theequation}{S\arabic{equation}} \setcounter{equation}{0}
\renewcommand{\thefigure}{S\arabic{figure}} \setcounter{figure}{0}

This Supplemental Material provides details about the CheMPS method, LCAO results with different coefficients, LDOS distributions under various conditions, analysis of in-gap states in the one-hole doped case, and results of the hybrid CPT+CheMPS approach.

\section{A. Chebyshev Matrix Product State Method\label{CheMPS}}

The quantity we aim to calculate is the lattice Green's function $G_{ij,\sigma}(z)=G^+_{ij,\sigma}(z)-G^-_{ij,\sigma}(z)$.
\begin{equation}
\begin{aligned}
&G_{ij,\sigma}^+=\langle \psi_0|c_{i\sigma} \frac{1}{z+(E_0-H)}c_{j\sigma}^\dagger |\psi_0\rangle,\\
&G_{ij,\sigma}^- = \langle \psi_0|c_{j\sigma}^\dagger \frac{1}{-z+(E_0-H)}c_{i\sigma}|\psi_0\rangle.
\end{aligned}
\end{equation}
Here, $|\psi_0\rangle$ is the ground state with energy $E_0$. 
The retarded Green's function can be obtained by setting $z=\omega+i\eta$ in $G_{ij,\sigma}(z)$. 
We implement the CheMPS method~\cite{Che1,Che2,Che3} to calculate the lattice Green's function. 
In the following, we illustrate how to compute $G_{ij,\sigma}^+(z)$. 
First, we choose an appropriate energy window $[-\omega_1,\omega_2]$ and perform a rescaling transformation for the variable $z$ and the Hamiltonian $H$.
\begin{equation}
\begin{aligned}
&z\mapsto z' = \omega'+i\eta/a, \  \omega'=\frac{\omega}{a}+b,\\
&H\mapsto H'=\frac{H-E_0}{a}+b,
\end{aligned}
\end{equation}
where $a=(\omega_1+\omega_2)/2W'$ and $b=(\omega_2-\omega_1)W'/(\omega_2+\omega_1)$. 
A small factor $W'=0.9875$ is introduced to make the algorithm perform better. 
The rescaled resolvent can be expanded by Chebyshev polynomials
\begin{equation}
\frac{1}{\pm z'-H'}=\sum_{n}\alpha_n^{\pm}(z')T_n(H'),
\end{equation}
where
\begin{equation}
\alpha_n^{\pm}(z)=\frac{2-\delta_{n,0}}{(\pm z)^{n+1}(1+\sqrt{z^2}\frac{\sqrt{z^2-1}}{z^2})^n\sqrt{1-\frac{1}{z^2}}}.
\end{equation}
The Green's function $G_{ij,\sigma}^+$ can be written as
\begin{equation}
G^+_{ij,\sigma}(z')=\frac{1}{a}\sum_{n=0}^{N_C-1}g_n^J\alpha_n^+(z')\mu_n^+,
\end{equation}
where $N_c$ is the cutoff of the polynomial order, and $g_n^J$ is the Jackson damping factor for suppressing Gibbs oscillations~\cite{kpm}.
The Chebyshev moments $\mu_n^+$ are given by
\begin{equation}
\begin{aligned}
\mu_n^+ = \langle \psi_0|c_{i\sigma}T_n(H')c_{j\sigma}^\dagger|\psi_0\rangle=\langle \psi_0|c_{i\sigma}|t_n\rangle.
\end{aligned}
\end{equation}
The Chebyshev vector $|t_n\rangle$ can be computed iteratively
\begin{equation}
\begin{aligned}
&|t_0\rangle = c_{i\sigma}^\dagger|\psi_0\rangle, |t_1\rangle = H'|t_0\rangle,\\
&|t_{n+1}\rangle = 2H'|t_n\rangle - |t_{n-1}\rangle.
\end{aligned}
\end{equation}
During calculation, we reduce the bond dimension of MPS using the variational compression method~\cite{Che1,SCHOLLWOCK}. 
The ground state is obtained with a bond dimension $D=2000$, and then compressed to $D_c=200$ during the Chebyshev iteration. 
The results remain qualitatively unchanged with larger bond dimensions $D=2500$ and $D_c=250$. 
The number of Chebyshev moments is cutoff as $N_c=500$, which is large enough to clearly show the DOS structure.
For $G_{ij,\sigma}^-(z)$, it is similar to computing $\mu^-_n=\langle \psi_0|c_{j\sigma}^\dagger T_n(H')c_{i\sigma}|\psi_0\rangle$. 
Putting them together, we finally obtain the lattice Green's function
\begin{equation}
G_{ij,\sigma}(z')=\frac{1}{a}\sum_{n=0}^{N_C-1}g_n^J(\alpha_n^+(z')\mu_n^+-\alpha_n^-(z')\mu_n^-).
\end{equation}

\section{B. More results of LCAO}
We present two different LCAO results in Fig.~1 and Fig.~3 of the main text, which closely resemble both experimental observations and our many-body numerical simulations. 
The guiding principles for selecting these LCAO coefficients are based on the symmetries of Wannier functions and molecular orbitals~\cite{hoffmann,hoffmannrmp}. 
Within this framework, the most critical factor is the sign structure of the LCAO coefficients, whereas their specific values are less important. 
This appendix presents various alternative sets of LCAO coefficients along with their corresponding patterns.

\begin{figure*}[tbp]
    \includegraphics[width=\linewidth]{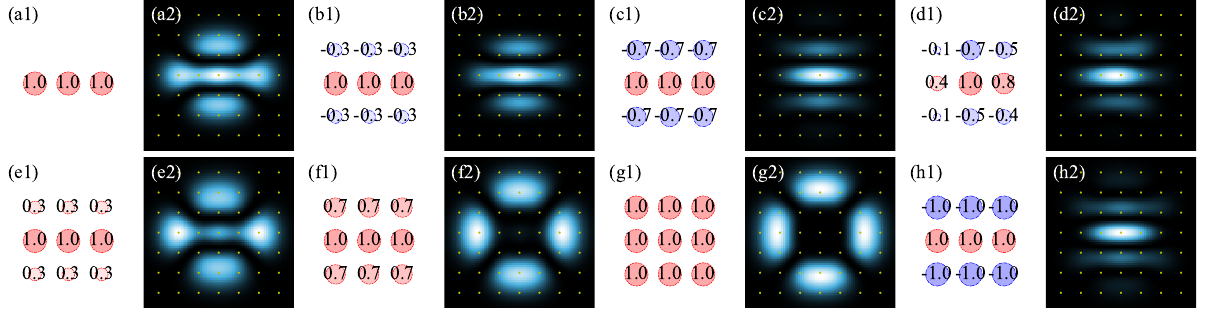}
    \caption{
    Different choices of LCAO coefficients and the corresponding molecular orbital density distributions.
    }\label{fig:aplcao}
\end{figure*}

First, the in-phase superposition of three atomic orbitals leads to a three-stripe structure, as shown in Fig.~\ref{fig:aplcao}(a1-a2). 
However, the side stripes are shorter than the central one. 
In Fig.~\ref{fig:aplcao}(b1-b2), we introduce six additional atomic orbitals on the sides, which possess opposite signs relative to the central three. 
This modification results in an extension of the side stripes. 
A similar three-stripe pattern is observed in Fig.~\ref{fig:aplcao}(c1-c2) and Fig.~\ref{fig:aplcao}(h1-h2).
Even when the LCAO coefficients are highly irregular but maintain the same sign structure, as seen in Fig.~\ref{fig:aplcao}(d1-d2), the resulting pattern remains consistent with experimental features. 
In contrast, when the difference in sign is removed, the side stripes become rounded and the ends of the central stripe appear broadened (Fig.~\ref{fig:aplcao}(e1-e2)).
Furthermore, when the correct sign structure is not maintained, the three-stripe structure is completely replaced by a larger clover pattern, as observed in Fig.~\ref{fig:aplcao}(f1-f2) and Fig.~\ref{fig:aplcao}(g1-g2). 
These findings underscore the importance of maintaining the correct sign structure in the LCAO coefficients to faithfully reproduce the symmetry properties of the molecular orbitals and achieve consistency with the experimental results.

\section{C. More Results for the one-hole case\label{ap1h}}
\begin{figure*}[tbp]
    \includegraphics[width=0.9\linewidth]{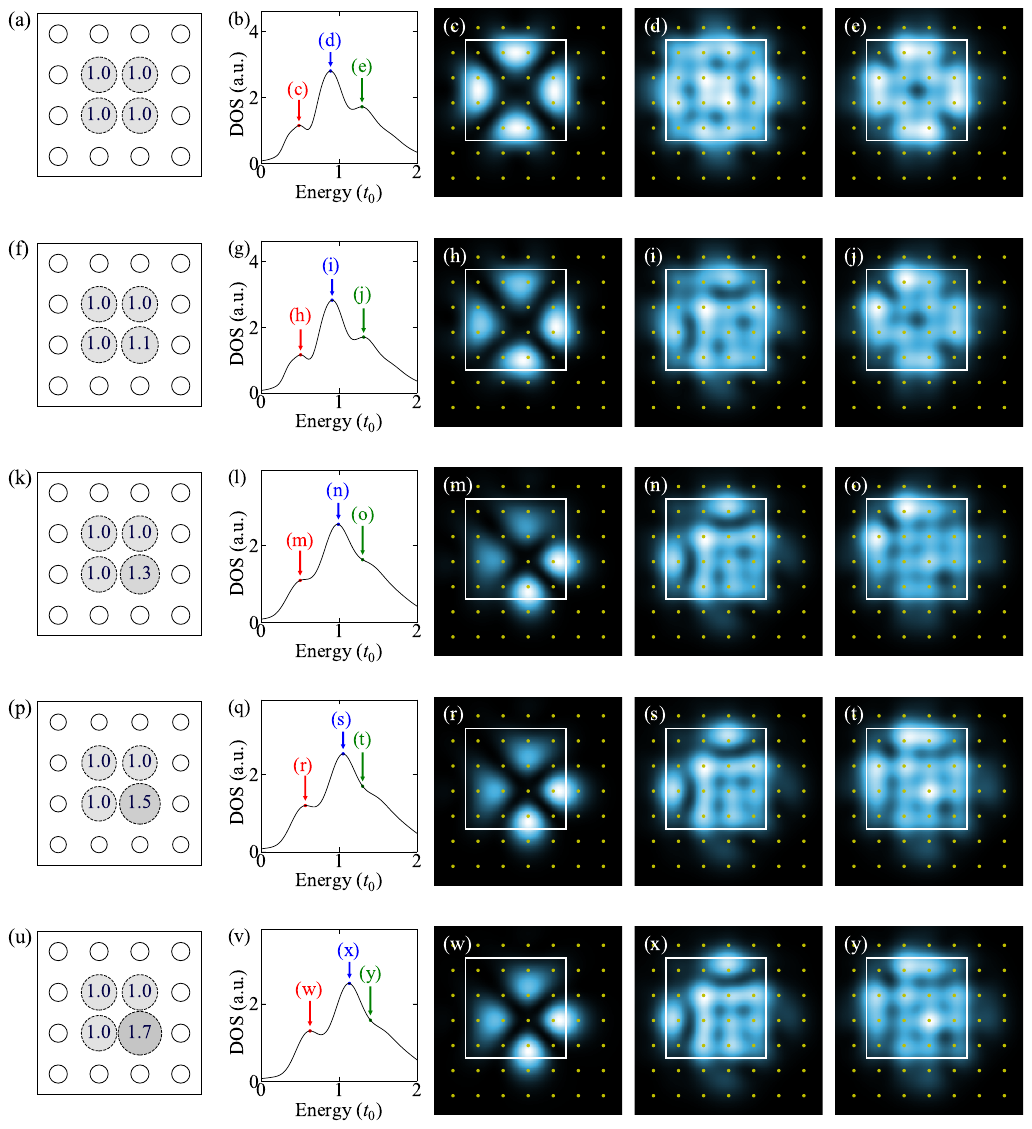}
    \caption{ 
    LDOS patterns for one-hole cases under various impurity potentials.
    For all cases, the chemical potential is $\mu=-3.7t_0$.
    }\label{fig:ap1h}
\end{figure*}

In the one-hole scenario, we observe that the four-lobe clover pattern, as seen in the experiment, has its center at the Cu site rather than the Na-doped site. 
Our analysis suggests that nearby impurities may break the symmetry of the potential fields originally induced by Na doping, resulting in a maximum potential field at a particular site. 
To further investigate this phenomenon and understand how the LDOS patterns evolve with the potential distribution, we examine more one-hole cases.
We start by considering a symmetric configuration, as shown in Fig.~\ref{fig:ap1h}(a-e). 
In this setup, the lower-energy state indicated by the red arrow suggests that a single-hole state influenced solely by the potential fields centered on the Na site would manifest itself as a linear superposition of four-lobe clovers on the surrounding Cu sites.
This would result in a larger four-lobe clover centered on the Na site. 
However, as we vary the potential field at the lower right lattice site (see Fig.~1(f-y)), we observe that the larger clover gradually shifts toward the lower right corner and eventually transforms into a configuration similar to the four-lobe clover centered on the Cu site. 
This evolution is consistent across different higher-energy states, indicating that it is related to changes in the potential fields.

\section{D. More Results for the two-hole case\label{ap2h}}
\begin{figure*}[tbp]
    \includegraphics[width=0.75\linewidth]{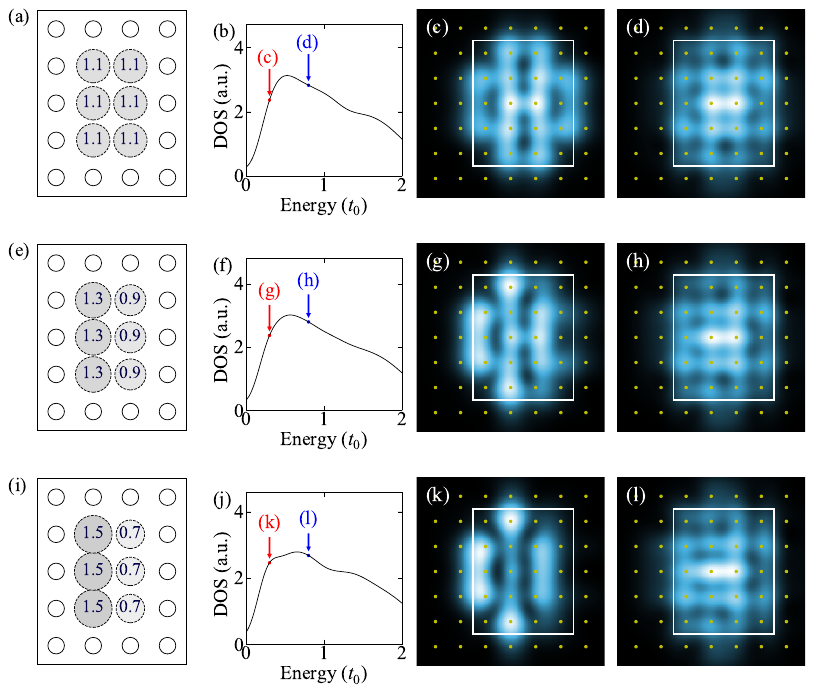}
    \caption{
    LDOS patterns for two-hole cases under various impurity potentials.
    For all cases, the chemical potential is $\mu=-3.7t_0$.
    }\label{fig:ap2h}
\end{figure*}

In addition to understanding the one-hole scenario, we are interested in exploring the behavior of two-holes within the lattice. 
Specifically, our aim is to investigate how local potential fields influence the resulting LDOS patterns.
As shown in Fig.~\ref{fig:ap2h}(a-d), when the local potential fields are uniform, the LDOS exhibits multiple intersecting stripes rather than a three-stripe structure.  
However, as we introduce an asymmetry between two columns of potentials (Fig.~\ref{fig:ap2h}(e, i)), the stripe pattern starts to emerge (Fig.~\ref{fig:ap2h}(g, k)).
At the same time, the ladder-shaped pattern becomes more pronounced (Fig.~\ref{fig:ap2h}(h, l)). 
These findings suggest that a two-hole state would not manifest itself as a three-stripe pattern if the potential field is completely uniform but would instead appear with an axial bias. 
This axial bias may arise from impurities or correlation effects among multiple holes, which can alter the effective local potential fields and the resulting LDOS patterns.

\section{E. More Results for the four-hole case\label{ap4h}}
\begin{figure*}[tbp]
    \includegraphics[width=\linewidth]{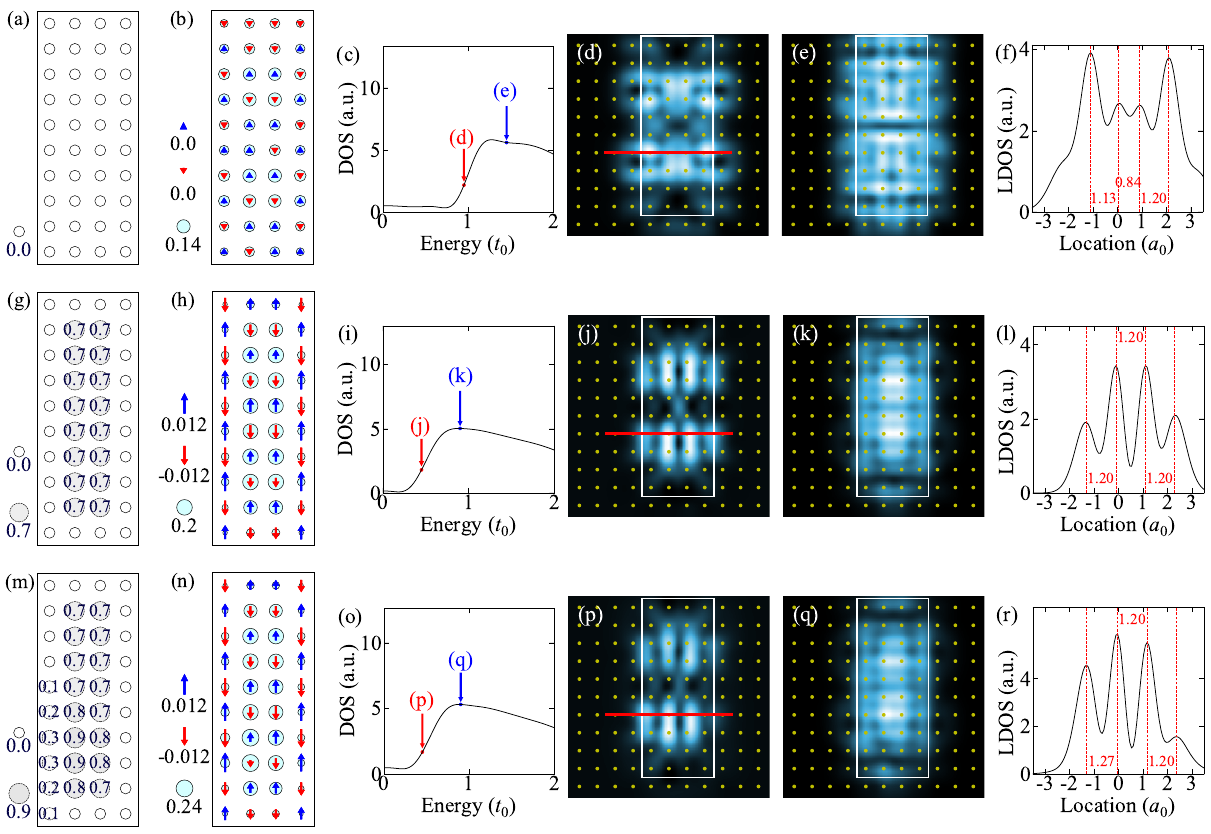}
    \caption{
    LDOS patterns for four-hole cases under various impurity potentials.
    For (a-f), the chemical potential is $\mu=-4.5t_0$. 
    For (g-l) and (m-r), the chemical potential is $\mu=-3.7t_0$.
    }
    \label{fig:ap4h}
\end{figure*}

We conclude our investigation by exploring the behavior of four holes within the lattice.
Our analysis begins with the scenario where there are no impurity potential fields, as depicted in Fig.~\ref{fig:ap4h}(a-e). 
In this case, the ground state exhibits a uniform hole distribution without a magnetic moment. 
However, the lower-energy LDOS (Fig.~\ref{fig:ap4h}(d)) still displays two distinct plaquettes. 
Upon closer inspection, we find that the hole distribution within these plaquettes does not show a striped pattern. 
As shown in Fig.~\ref{fig:ap4h}(f), the LDOS distribution along the red cut line in Fig.~\ref{fig:ap4h}(d) exhibits four peaks with intervals of $1.13a_0$, $0.84a_0$, and $1.20a_0$ from left to right.
The side peaks are taller than the middle ones. 
When a uniform potential field is introduced (Fig.~\ref{fig:ap4h}(g-l)), we observe that the stripe pattern becomes evident and the stripe order emerges in the ground state. 
This indicates that the impurity potential can stabilize the stripe order and the corresponding LDOS pattern. 
As depicted in Fig.~\ref{fig:ap4h}(l), the LDOS distribution along the red cut line in Fig.~\ref{fig:ap4h}(j) still shows four peaks, but now the middle two are the tallest, with peak spacings of $1.20a_0$. 
Finally, we modulate the potential field to be stronger on one side (Fig.~\ref{fig:ap4h}(m-r)), resulting in only three stripes being evident in Fig.~\ref{fig:ap4h}(r).
The rightmost peak is significantly lower than the others, with distances between neighboring peaks ranging from $1.2a_0$ to $1.3a_0$, consistent with experimental observations.
These results further emphasize the significance of impurity potentials and correlation effects between multiple holes.

\section{F. The split of the two in-gap peaks in the one-hole doped case}

In the main text, the emergence of peaks for in-gap states demonstrates the phenomenon of spectral weight transfer~\cite{RUAN20161826,cai2016visualizing}. 
To further explore the role of underlying antiferromagnetic correlations, we investigate their contribution to the peak splitting. 
Starting from the spectral function formalism~\cite{swt}, the spectral weight associated with a particular eigenstate is directly related to the overlap coefficients
\begin{equation}
\left|\langle \psi_m^{N}|c_{i\sigma}^\dagger|\psi_0^{N-1}\rangle\right|^2\delta(\omega-E_{m}^N+E_0^{N-1}),
\end{equation}
where $|\psi_m^N\rangle$ is the $m$-th eigenstate in the $N$-electron subspace (half-filling), and $|\psi_0^{N-1}\rangle$ is the ground state in the $(N-1)$-electron subspace (one-hole doping). 
In this context, the positions of the peaks reflect the eigenenergies of the half-filling system.

\begin{figure*}[tbp]
    \includegraphics[width=0.7\linewidth]{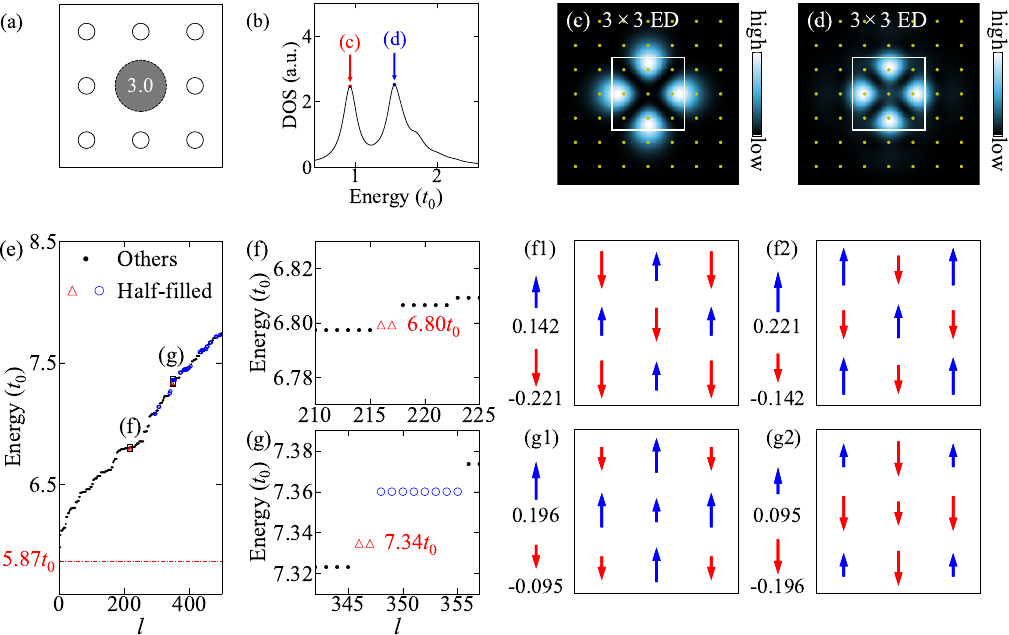}
    \caption{
    (a-d) Single-hole-doped results for the $3\times3$ cluster with an impurity potential located at its center.
    (e) The $500$ lowest eigenenergies with their order labeled as $l$.
    Hollow markers denote the eigenstates in the half-filling subspace.
    (f-g) The enlarged views of (e).
    (f1-f2) Spin configurations for the two eigenstates marked by red triangles in (f). 
    (g1-g2) Spin configurations for the two eigenstates marked by red triangles in (g).
    }
    \label{fig:apF0}
\end{figure*}

To identify the wavefunctions associated with the two in-gap peaks, we perform ED calculations on a $3\times3$ cluster subject to the potential shown in Fig.~\ref{fig:apF0}(a), keeping all other parameters identical to those in the single-hole-doped example presented in the main text (Fig.~2(a)). 
As illustrated in Fig.~\ref{fig:apF0}(b), the spectral function exhibits two prominent peaks at energies $0.93t_0$ and $1.47t_0$. 
The corresponding LDOS patterns shown in Figs.~\ref{fig:apF0}(c-d) closely resemble those in the $4\times4$ lattice (Fig.~2(c-d)).

To understand the nature of these two in-gap states, we calculate the lowest $500$ eigenenergies for the system (Fig.~\ref{fig:apF0}(e)), with the hollow markers representing the energies in the half-filling subspace. 
Specifically, we focus on four half-filling eigenstates near the characteristic in-gap peaks, highlighted by red triangles in Fig.~\ref{fig:apF0}(f-g).
The two eigenstates in Fig.~\ref{fig:apF0}(f) are the ground states of the half-filling subspace—degenerate due to the odd number of lattice sites~\cite{LiebTheorem}—and have energy differences of $6.80t_0-5.87t_0=0.93t_0$ relative to the single-hole-doped ground state (dashed line in Fig.~\ref{fig:apF0}(e)), consistent with the lower-energy peak in Fig.~\ref{fig:apF0}(b).
Magnetic distribution analysis of these states reveals a clear antiferromagnetic spin alignment, as shown in Fig.~\ref{fig:apF0}(f1–f2).
In addition, we identify two degenerate excited states in Fig.~\ref{fig:apF0}(g), whose energy difference relative to the single-hole-doped ground state is $7.34t_0-5.87t_0=1.47t_0$, aligning with the position of the higher-energy peak. 
These states feature a flipped central spin embedded in an antiferromagnetic background, as illustrated in Fig.~\ref{fig:apF0}(g1–g2). 
Hence, we identify the above states as the dominant contributors to the two in-gap peaks.

\begin{figure*}[tbp]
    \includegraphics[width=0.7\linewidth]{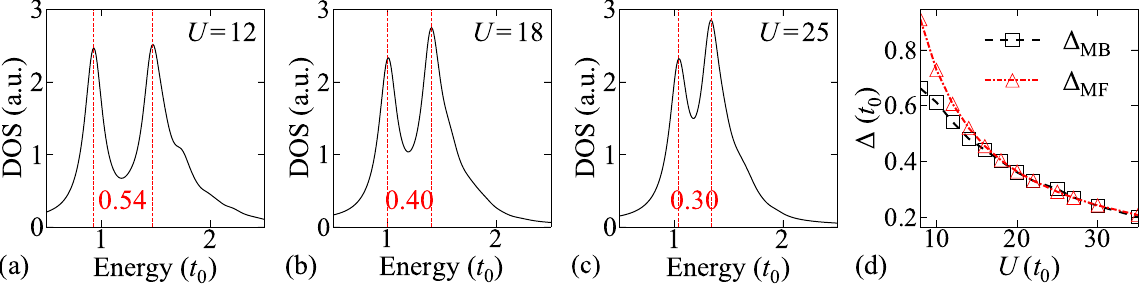}
    \caption{
    (a-c) Two in-gap peaks in the DOS for the one-hole doped case under various $U$. 
    (d) Comparison of the energy splitting between numerical results and mean-field estimations.
    }
    \label{fig:apF2}
\end{figure*}

These findings allow us to interpret the energy splitting between the peaks using an effective spin model that incorporates antiferromagnetic couplings between nearest and next-nearest neighbors
\begin{equation}
    H = J_0\sum_{\langle ij\rangle}\bm{S}_i\cdot \bm{S}_j + J_1\sum_{\langle\langle ij\rangle\rangle}\bm{S}_i\cdot \bm{S}_j,
\end{equation}
where $J_0=4t_0^2/U$ and $J_1=4t_1^2/U$. 
By analyzing this model, we obtain a mean field (MF) estimation of the peak splitting as $\Delta_{\rm{MF}}=2(J_0-J_1)$. 
When the Hubbard interaction $U$ increases, both effective spin couplings $J_0$ and $J_1$ decrease, leading to a reduction in peak splitting, as shown in Fig.~\ref{fig:apF2}(a–c). 
Furthermore, Fig.~\ref{fig:apF2}(d) compares the many-body (MB) numerical results of the original Hubbard model $\Delta_{\textrm{MB}}$ with the MF results of the effective spin model $\Delta_{\textrm{MF}}$. 
The agreement between these two results in the strong-coupling regime confirms that the observed peak splitting is primarily governed by effective antiferromagnetic interactions.

\begin{figure*}[tbp]
    \includegraphics[width=1.01\linewidth]{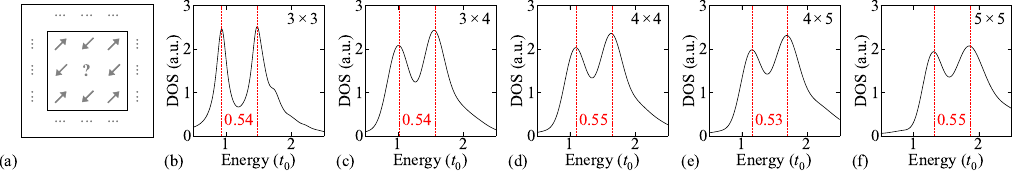}
    \caption{
    (a) The two in-gap states mainly originate from the local properties of the $3 \times 3$ cluster with a central potential. 
    (b-f) In-gap states for the one-hole doped systems with various cluster sizes and $U=12t_0$.
    }
    \label{fig:apF1}
\end{figure*}

For lattices larger than $4\times4$, exponential growth in computational complexity makes it impractical to access full spectrum and excited-state wave functions directly.
However, we note that doped holes remain confined by local potentials.
The eigenstates that contribute dominantly to the two in-gap spectral peaks exhibit inherent spatial locality, and the spin configuration illustrated in Fig.~\ref{fig:apF1}(a) remains a reliable description.
We therefore expect that the splitting energies will be primarily determined by localized spin interactions within the $3\times3$ region.
This expectation is further verified by the spectral functions presented for various lattice sizes in Fig.~\ref{fig:apF1}(b-f). 
Specifically, all cluster geometries from $3\times3$ to $5\times5$ display two in-gap peaks, with little size dependence observed in the energy difference between peaks, fluctuating only within a narrow range of $0.53t_0–0.55t_0$. 
This quantitative insensitivity confirms the local nature of the in-gap states and suggests that the peak splitting arises from the local interactions between electrons and the antiferromagnetic background.

\section{G. Cluster Perturbation Theory Analysis}
In the main text, we calculate single-particle Green’s functions on finite-size clusters.
In this appendix, we further test the robustness of our findings by considering larger systems using cluster perturbation theory (CPT). 

Experimentally, holes are introduced by substituting Ca with Na atoms, which unavoidably generate impurity potentials.  
In our simulations, these potentials are assumed to induce a localized charge density distribution, making finite-cluster simulations sufficient to capture the essential real-space electronic structure. 
To verify the validity of this assumption, we employ CPT to effectively enlarge the cluster size and investigate how it impacts the real-space structure of the in-gap states. 
If this structure remains unchanged as the cluster size increases, it confirms that our main conclusions are robust and not artifacts of finite-size effects.

We begin by reviewing the basic framework of CPT. 
In the lattice representation, the single-particle Green’s function matrix is given by
\begin{equation}
\mathbf{G}(z)=\frac{1}{z-\mathbf{t}-\mathbf{\Sigma}(z)},
\end{equation}
where $\mathbf{t}$ denotes the hopping matrix and $\mathbf{\Sigma}(z)$ the electronic self-energy matrix. 
In the CPT approach, the system is partitioned into smaller clusters $\{\bm{R}\}$. 
The Hamiltonian is decomposed as
\begin{equation}
H=H’+V=\sum_{\bm{R}} H^{(\bm{R})}+\sum_{\alpha\beta}(\mathbf{t}_{ic})_{\alpha\beta} c^\dagger_{\alpha}c_\beta,
\end{equation}
where $(\mathbf{t}_{ic})_{\alpha\beta}$ represent inter-cluster hoppings, including both nearest- and next-nearest-neighbor terms. 
Under the CPT approximation, the self-energy matrix is assumed to take a block-diagonal form with respect to clusters~\cite{Pavarini_2015,CPT1,CPT2,realspaceCPT,Lijianxin,zhao2024chebyshevpseudositematrixproduct}
\begin{equation}
\mathbf{\Sigma}(z)\approx \bigoplus_{\bm{R}} \mathbf{\Sigma}^{(\bm{R})}(z),
\end{equation}
with $\mathbf{\Sigma}^{(\bm{R})}(z)$ denoting the self-energy matrix of cluster $\bm{R}$. 
This yields an approximate expression for the Green’s function of the full system,
\begin{equation}
\mathbf{G}^{-1}(z)\approx \bigoplus_{\bm{R}} \mathbf{G}^{(\bm{R})}(z)^{-1} - \mathbf{t}_{ic},
\label{eq:CPT}
\end{equation}
where $\mathbf{G}^{(\bm{R})}(z)$ is the Green’s function matrix of cluster $\bm{R}$.

\begin{figure*}[tbp]
    \includegraphics[width=\linewidth]{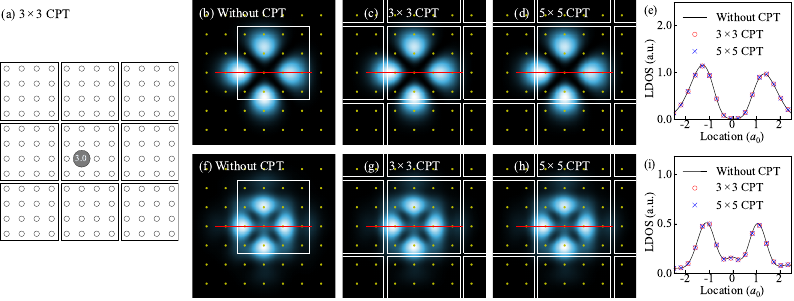}
    \caption{
    (a) Illustration of the $3 \times 3$ supercluster used for CPT calculation in the one-hole doped case. 
    (b-e) Comparison of LDOS results for the low-energy state at $\omega=1.08t_0$. 
    (b) Without CPT. 
    (c) $3\times 3$ supercluster using CPT. (d) $5\times 5$ supercluster using CPT. 
    (f-i) Comparison of LDOS results for the high-energy state at $\omega=1.63t_0$.
    (f) Without CPT.
    (g) $3\times 3$ supercluster using CPT. 
    (h) $5\times 5$ supercluster using CPT. 
    }
    \label{fig:apG1}
\end{figure*}

\begin{figure*}[tbp]
    \includegraphics[width=\linewidth]{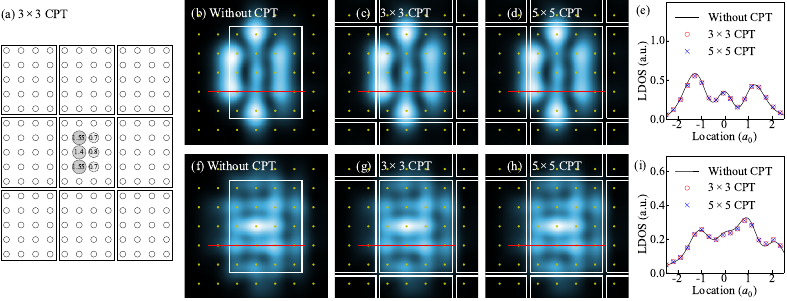}
    \caption{
    (a) Illustration of the $3 \times 3$ supercluster used for CPT calculation in the two-hole doped case. 
    (b-e) Comparison of LDOS results for the low-energy state at $\omega=0.32t_0$.
    (b) Without CPT.
    (c) $3\times 3$ supercluster using CPT.
    (d) $5\times 5$ supercluster using CPT.
    (f-i) Comparison of LDOS results for the high-energy state at $\omega=0.8t_0$.
    (f) Without CPT.
    (g) $3\times 3$ supercluster using CPT.
    (h) $5\times 5$ supercluster using CPT. 
    }
    \label{fig:apG2}
\end{figure*}

\begin{figure*}[tbp]
    \includegraphics[width=0.94\linewidth]{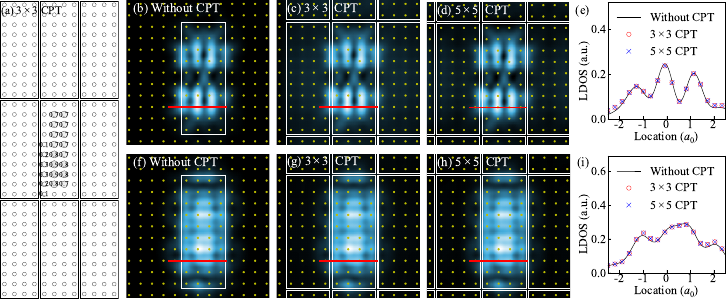}
    \caption{
    (a) Illustration of the $3 \times 3$ supercluster used for CPT calculation in the four-hole doped case. 
    (b-e) Comparison of LDOS results for the low-energy state at $\omega=0.5t_0$.
    (b) Without CPT.
    (c) $3\times 3$ supercluster using CPT.
    (d) $5\times 5$ supercluster using CPT.
    (f-i) Comparison of LDOS results for the high-energy state at $\omega = 0.9t_0$.
    (f) Without CPT.
    (g) $3\times 3$ supercluster using CPT.
    (h) $5\times 5$ supercluster using CPT.
    }
    \label{fig:apG3}
\end{figure*}

In our simulations, we construct larger superclusters by assembling multiple geometrically identical clusters. 
As illustrated in Figs.~\ref{fig:apG1}(a), \ref{fig:apG2}(a), and \ref{fig:apG3}(a), a $3\times 3$ arrangement of clusters forms a $3\times 3$ supercluster. 
The impurity cluster studied in the main text is embedded at the central position $\bm{R}_0$ of the supercluster, while the surrounding positions are occupied by environmental clusters $\{\bm{R}’\}$. 
Each environmental cluster has a Hamiltonian given by
\begin{equation}
\begin{aligned}
H^{(\bm{R'})}=-t_0\sum_{\langle ij\rangle,\sigma}c_{i\sigma}^\dagger c_{j\sigma} -t_1\sum_{\langle \langle ij\rangle \rangle,\sigma}c_{i\sigma}^\dagger c_{j\sigma}-\mu\sum_i n_i+U\sum_i(n_{i\uparrow}-\frac{1}{2})(n_{i\downarrow}-\frac{1}{2}),
\end{aligned}
\end{equation}
where the parameters $t_0=1.0$, $t_1=-0.3t_0$, $U=12t_0$, and $\mu=-3.7t_0$ are identical to those of cluster $\bm{R}_0$ in the main text. 
Owing to the absence of impurity potential terms, the environmental clusters remain half-filled in the ground state.

To compute the lattice Green’s functions $\mathbf{G}^{(\bm{R})}(z)$ of the clusters, we use the CheMPS method and subsequently couple them through the CPT formalism (Eq.~(\ref{eq:CPT})). 
We then transform the resulting Green’s functions into the real-space representation, from which we obtain the LDOS $\rho(\bm{r},\omega)$. 
Following the same procedure, we also perform calculations on $5\times 5$ superclusters to investigate the effects of larger sizes. 
The results for the one-hole, two-hole, and four-hole doping cases are shown in Figs.~\ref{fig:apG1}, \ref{fig:apG2}, and \ref{fig:apG3}, respectively.
The bond dimension used in the ground state calculation is set to $D=2000$, which is then compressed to $D_c = 200$ during the Chebyshev iterations. 
The number of Chebyshev moments is cutoff as $N_c = 500$.

For the one-hole doped case, the $4\times 4$ impurity cluster $\bm{R}_0$ is initially computed without CPT. 
After that, we calculate approximate Green’s functions for $\bm{R}_0$ embedded in both $3\times 3$ and $5\times 5$ supercluster environments using CPT. 
As depicted in Figs.~\ref{fig:apG1}(b–d), the LDOS distributions at the low-energy state $\omega=1.08t_0$ exhibit excellent agreement across these three cases. 
Furthermore, we analyze consistency by taking a line cut in real space (red line) and comparing the LDOS across all cases. 
As shown in Fig.~\ref{fig:apG1}(e), introducing environmental clusters in the $3\times 3$ and $5 \times 5$ CPT calculation produces no noticeable differences in the central impurity cluster $\bm{R}_0$. 
Similarly, comparisons for the high-energy state at $\omega=1.63t_0$ in Figs.~\ref{fig:apG1}(f–i) confirm that all three cases yield consistent results.

We further expand our systematic comparison to two-hole and four-hole doped systems. 
For the two-hole case, a $5\times 4$ impurity cluster $\bm{R}_0$ is embedded in both $3\times 3$ and $5\times 5$ supercluster environments. 
As shown in Figs.~\ref{fig:apG2}(b–e), the LDOS distributions for the low-energy state at $\omega=0.32t_0$ exhibit excellent agreement across three cases.
Moreover, the high-energy results at $\omega=0.8t_0$ in Figs.~\ref{fig:apG2}(f–g) also demonstrate consistent outcomes.

For the four-hole doped system with a $10\times 4$ impurity cluster $\bm{R}_0$, the same CPT scheme is applied to obtain Green’s functions in both $3\times 3$ and $5\times 5$ supercluster environments. 
The LDOS distributions at low energy ($\omega=0.5t_0$) and high energy ($\omega=0.9t_0$) are presented in Figs.~\ref{fig:apG3}(b–e) and Figs.~\ref{fig:apG3}(f-g), respectively. 
The results consistently indicate that the environmental coupling through CPT does not have a significant impact on the electronic structure of $\bm{R}_0$.
In particular, all schemes reproduce (i) intrinsic stripe patterns with aligned peak positions at low energy, and (ii) robust ladder patterns at high energy. 
Moreover, the LDOS shows clear convergence between the $3\times 3$ and $5\times 5$ CPT environments.

In summary, by combining CheMPS with CPT, we successfully extend our analysis to larger systems. 
The results confirm that the key conclusions presented in the main text remain unchanged, demonstrating the robustness of the physical phenomena observed in finite clusters and ensuring that they are not artifacts of finite-size effects.

\end{document}